\documentclass[structabstract]{raa}
\usepackage{booktabs}
\usepackage{graphicx,times}
\usepackage{natbib,amssymb}
\usepackage{epsfig}
\usepackage{float}
\usepackage{amsmath}
\usepackage{ulem}

\usepackage{mathtools}

\providecommand{\bm}[1]{\mbox{\boldmath$#1$\unboldmath}}

\begin{document}

  \title{The fractions of post-binary-interaction stars and evolved blue straggler stars on the red giant branch of globular clusters}

   \volnopage{ {\bf 20XX} Vol.\ {\bf X} No. {\bf XX}, 000--000}
   \setcounter{page}{1} 
   
%% AUTHOR/INSTITUTIONS FOR AASTEX6.1:
\author{Dandan~Wei\inst{1,2,3,4}
        \and Bo~Wang\inst{1,2,4}
        \and Hailiang~Chen,\inst{1,2,4}
        \and Haifeng~Wang\inst{5} \footnote{LAMOST fellow }
        \and Xiaobo~Gong\inst{1,2,3,4}
        \and Dongdong Liu \inst{1,2,4}
        %\and David~Katz\inst{9}
        \and Dengkai~Jiang\inst{1,2,4}
        }
        
\institute{Yunnan Observatories, Chinese Academy of Sciences (CAS), 396 Yangfangwang, Guandu District, Kunming 650216, P.R. China; 
          {\it  wdd@ynao.ac.cn, dengkai@ynao.ac.cn
           }\\
           \and
           Key Laboratory for the Structure and Evolution of Celestial Objects, CAS, Kunming 650216, P.R. China
           \and
           University of Chinese Academy of Science, Beĳing 100049, P.R. China
           \and
           Center for Astronomical Mega-Science, Chinese Academy of Sciences, Beijing 100012, P.R. China
           \and
           South-Western Institute for Astronomy Research, Yunnan University, Kunming 650500, P.R. China    
              }

\abstract
{The red giant branch (RGB) of globular clusters (GCs) is home to some exotic stars, which may provide clues on the formation of multiple stellar populations in GCs. It is well known that binary interactions are responsible for many exotic stars. Thus, it is important to understand what fraction of stars on the RGB of GCs is the result of binary interactions. In this paper, we performed a binary population synthesis study to track the number of post-binary-interaction (post-BI) stars that appear on the RGB, with particular emphasis on the evolved blue straggler stars (E-BSSs). Assuming an initial binary fraction of nearly 50\%, we find that about half of the objects on the RGB (called giants) underwent the binary interactions, and that E-BSSs account for around 10\% of the giants in our standard simulation. We also compare the properties of post-BI giants that evolved from different channels. We find that the initial orbital period and mass ratio distributions significantly affect the fraction of post-BI giants. Our results imply that the non-standard stars from binary interactions provide a non-negligible contribution to the RGB stars in GCs, which should be considered in future investigations of the origin of multiple stellar populations.
}

   \authorrunning{Wei et al. }            %author_head in even pages
   \titlerunning{The fractions of post-BI giants and E-BSSs}  % title_head in odd pages
   \maketitle

%Hurley's rapid binary evolution code in conjunction with a series of Monte Carlo simulations 
%Most of giants have experienced wind accretion channel at some stages of their evolution, while only a small fraction formed from other channels. 

\section{INTRODUCTION}
\label{introduce}
The chemical inhomogeneities among stars in Galactic globular clusters (hereafter GCs) have been explored for many years, called multiple stellar populations \citep{2012A&ARvGratton,2018ARA&ABastian}. Large star-to-star abundance variations have been found for light elements (such as C, N, O, Na and Al) in many GCs, and the most outstanding one is Na-O anti-correlation \citep{2009bA&ACarretta,2009aA&ACarretta}. Stars in GCs may be roughly classified into two populations: normal population and chemically peculiar population. The former contains stars with O and Na abundances similar to the field stars, and the latter contains stars with abnormal abundances (such as enhancement in Na and N, and depletion in O and C). Nearly 50\% $\sim$ 70\% of stars in a GC belong to the chemically peculiar population \citep{2009bA&ACarretta,2009aA&ACarretta}. The chemical inhomogeneities were mainly found on the red giant branch (RGB) of GCs \citep{2009bA&ACarretta,2009aA&ACarretta}, and similar abundance variations were also found in some unevolved stars, e.g. main sequence turn-off (MSTO) stars and subgiant-branch stars.

Among the scenarios proposed to explain the multiple stellar populations, the most popular one is the self-enrichment scenario \citep{2004ARA&AGratton}, which includes more than one star-formation episodes. The second-generation stars with chemical inhomogeneities formed from the ejecta of the first-generation stars. Based on different origins of the ejecta, there are a variety of possible candidate polluters in the first-generation stars, e.g. fast-rotating massive stars \citep{2007A&ADecressin}, massive interacting binaries \citep{2009A&AdeMink}, intermediate-mass close binaries \citep{2012A&AVanbeveren}, supermassive stars \citep{2014MNRASDenissenkov} and asymptotic giant branch stars \citep{2009A&ADecressin,2010MNRASD'Ercole}. However, the self-enrichment scenario still has some problems \citep{2006A&APrantzos,2015MNRASRenzini}. For example, this scenario has difficulty in explaining such a large amount of abnormal stars observed, according to the initial mass function \citep{2018ARA&ABastian}. Therefore, it is likely that more than one scenario work together in generating such a large fraction of abnormal stars in GCs, e.g. RGB self-enrichment by extra-mixing \citep{1979ApJSweigart,1994PASPKraft}.

While the majority of observational and theoretical studies about multiple stellar populations are in the framework of single star scenario, some RGB stars observed in GCs may not be normal single stars (or "true single stars"). They might be post-binary-interaction (post-BI) stars that experienced mass transfer, merger or wind accretion. In general, the binaries are expected to play an important role during the evolution of GCs \citep{2019A&ARvGratton}. Binary interactions can produce peculiar objects in GCs, such as cataclysmic variables \citep{2018MNRASSandoval}, X-ray binaries \citep{2020ApJSLehmer}, compact intermediate-mass black-hole X-ray binaries \citep{2020ApJChen}, millisecond pulsars \citep{2020ApJDai, 2020ApJPan}, blue straggler stars \citep{2020ApJSingh} and so on. In addition, such interactions can explain a part of stars with exotic abundances in GCs as well as in the field, e.g. barium stars \citep{1988A&ABoffin, 1995MNRASHana}, carbon-enhancement metal-poor stars \citep{2008A&ALugaro,2013A&AAbate}, fluorine-enrichment stars \citep{2008A&ALugaro}, etc. More importantly, \cite{2012A&ARvGratton} has suggested that the binary interactions may contribute to the alterations of the surface abundances of stars. For example, some binary interactions have been used to explain the chemically peculiar population in GCs, e.g. tidally enforced enhanced extra mixing \citep{2006ApJDenissenkov}, rotationally induced mixing \citep{2014ApJJiang}, and stable mass transfer \citep{2020MNRASWei}.

The fraction of post-BI stars on the RGB of GCs remains unclear at present, although this fraction may be one of the most important parameters affecting the theoretically predicted fraction of the chemically peculiar population. For example, the main-sequence (MS) stars, which have experienced binary interactions (e.g. merger, mass accretion through mass transfer or wind accretion), are expected to evolve to the RGB, and mix with normal single RGB stars in the color-magnitude diagram (CMD). Therefore, it is a quite hard task to distinguish post-BI stars from normal stars on the RGB of GCs.

Actually, some samples observed on the RGB have already been confirmed as binary systems \citep{1983ApJHarris}. Furthermore, a small number of observation studies have investigated the binary fraction in different populations of GCs, and they found that the fraction of binaries in the first-generation population is higher than that in the second-generation stars \citep{2015A&ALucatello,2018ApJDalessandro}.  But the fraction of post-BI population on the RGB is still unclear. In this paper, we will focus on the properties of post-BI stars on the RGB of GCs.

A possible kind of the post-BI stars on the RGB of GCs is evolved blue straggler stars (E-BSSs). BSSs are brighter than the MSTO stars of the host cluster \citep{1953AJSandage,1992MNRASFerraro,1993AJFerraro, 2008A&AMoretti,2011ApJBeccari,2012ApJBeccari,2016MNRASSimunovic}, which may be created by binary evolution \citep{1964MNRASMcCrea,2001AJCarney,2008MNRASChen} or stellar collision \citep{1976ApLHills,1995ARA&ABailyn}. BSSs will be called E-BSSs, when they evolve to the post-MS evolutionary phases (e.g. RGB). A few E-BSSs have been identified by masses which are higher than masses of other giants \citep{2016ApJFerraro,2018AJ.Li}. \citet{2019A&ARvGratton} has roughly estimated that E-BSSs account for about 6\% of giants. We will further study the fraction of E-BSSs in giants in a detailed binary population synthesis (BPS) study considering the binary interactions.

In this paper, we performed a detailed BPS study to investigate the characteristics of the post-BI stars and E-BSSs on the RGB. In Section \ref{methods}, we show the four channels of binary interactions and introduce the methods of computations. The results are presented in Section \ref{result}. Finally, we give the discussions and conclusions in Section \ref{discussion}.

\section{Methods}
\label{methods}{}

\begin{figure}
\begin{center}
\includegraphics[width=0.7\textwidth]{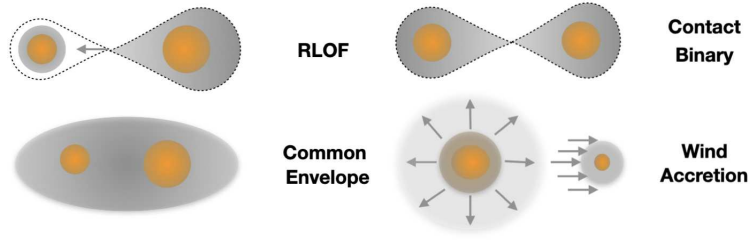}
\caption{Schematic illustration of the four channels of binary interactions: RLOF channel,  common envelope channel, contact binary channel and wind accretion channel. }
\label{fig:cartoon}
   \end{center}
\end{figure}

To investigate the properties of post-BI stars on the RGB, we performed a series of Monte Carlo simulations  by using the rapid binary-evolution algorithm from Hurley \citep{2000MNRASHurley,2002MNRASHurley}. In each of our simulations, we evolved $10^{6}$ systems, and a single starburst was assumed. The binary interactions in this paper include four channels: common envelope channel, Roche lobe overflow (RLOF) channel, contact binary channel and wind accretion channel, as exhibited in Fig.\,\ref{fig:cartoon}.

\subsection{Four channels of the binary interactions}

\subsubsection{Roche lobe overflow channel}
\label{RLOF}
The mass is transferred from the donor to the accretor, as long as the donor fills its Roche lobe. There are three types of mass transfer: nuclear time-scale, thermal time-scale, and dynamical time-scale \citep[for more details see][] {2000MNRASHurley,2002MNRASHurley}. 
The mass will be transferred from the donor on a dynamical time-scale, if the mass ratio  $q = M_{1}/M_{2}$ is larger than a critical mass ratio $q_{\rm c}$, where $M_{1}$  and $M_{2}$  are the masses of the donor and of the accretor, respectively. Different evolutionary stages of the donor at the onset of mass transfer have different critical mass ratios $q_{\rm c}$ \citep{1987ApJHjellming, 2002MNRASHan, 2002ApJPodsiadlowski, 2010ApJGe, 2020RAAHan}. In this paper, $ q_{\rm c} = 3.0 $ and $4.0$ when the donor is on the MS stage and Hertzsprung gap stage, respectively. If the primordial donor is on the first giant branch (GB) or asymptotic giant branch (AGB) stage, we use 
\begin{equation}
q_{\rm c} = {[1.67 - x + 2\,(\frac{M_{\rm c1}}{M_{1}})^5]}/{2.13},
\end{equation}
where $R_{1}\propto M_{1}^{-x}$ for normal giants, $M_{1}$ is the mass of the donor, and $M_{\rm c1}$ is its core mass. If the donor is a naked helium giant, $ q_{\rm c} = 0.748 $ \cite[for more details see][]{2002MNRASHurley}. In this paper, the RLOF channel includes the binary systems that experience RLOF, but do not undergo common envelopes or contact phase.

%$ x = {\rm d\ ln\ }R_{1}/{\rm d\ ln\ }M_{1} $ is the mass-radius exponent of the donor and varies with composition.

\subsubsection{Common envelope channel}

The case with  $q > q_{\rm c}$ in which the donor is a giant begins common-envelope evolution. The accretor and the core of the giant are all contained within the common envelope, when the system enters the common envelope. According to the \citet{1997MNRASTout}, the common envelope is treated as follows. The orbital energy will be transferred to the common envelope due to the frictional drag produced by the common envelope, when the accretor spirals in \citep{1988ApJLivio}. The entire common envelope is assumed to be removed away from the system, if 

\begin{equation}
-\frac{\rm G}{\lambda} \left(\frac{M_{1}^{\rm i}M_{\rm env1}^{\rm i}}{R_{1}} + \frac{M_{2}^{\rm i}M_{\rm env2}^{i}}{R_{2}} \right) =  \alpha_{\rm ce} \left( \frac{{\rm G}M_{1}^{\rm i}M_{2}^{\rm i} }{2a_{\rm i}} - \frac{ {\rm G} M_{\rm 1c}^{\rm f} M_{\rm 2c}^{\rm f} }{2a_{\rm f}}\right).
\label{CE}
\end{equation}

The left part is the bind energy of the common envelope, and the right part is the release of the orbital energy. The indices i and f are used to represent values before and after the common envelope, respectively. $M_{1}$ and $ M_{2}$ represent the masses of the donor and the accretor, respectively. The masses of the core of the two components are $ M_{\rm 1c} $ and $ M_{\rm 2c}$, respectively.  The mass of the donor's envelope is $M_{\rm env1} $, and that of the accretor is $ M_{\rm env2}$. $R_{1}$ and $R_{2}$ are the radii of the donor and the accretor, and $a$ is the orbital separation. $ \rm \lambda$ is a structure parameter which is related to the evolutionary stage of the donor, but it is still unclear about the calculation of $\rm \lambda$. Here we set $\rm \lambda = 0.50$ ($\rm \lambda$ = 0.25, 0.75, and 1.00 are discussed in Section\,\ref{discussion}). $\alpha_{\rm ce} $, common envelope ejection efficiency, represents the fraction of the released orbital energy used to eject the common envelope. We set $\alpha_{\rm ce} = 1$ in this paper, as suggested in the \citet{2002MNRASHurley}.

\subsubsection{Contact binary channel}
As a result of RLOF, the binary is also possible to evolve into contact state. The binary system may become a contact binary (e.g. W UMa-type binary systems), if the MS accretor fills its own Roche lobe due to mass accretion from the MS donor during nuclear or thermal time-scale mass transfer. Two MS components in a contact binary are assumed to coalesce in this work, and the merged star will evolve subsequently to the RGB. Besides, the total mass of the contact binary system is assumed to be conservative during merger process. It should be noted that a binary will be classified as the common envelope channel if this binary not only experiences contact state, but also undergoes the common envelope state.

\subsubsection{Wind accretion channel}

 When the donor evolves to the giant and beyond, the mass-loss rate of the donor is applied according to \citet{
1978A&AKudritzki}
\begin{equation}
\dot{M}_{\rm R} = 4\times 10^{-13}\,\frac{\eta\,L_{1}\,R_{1}}{M_{1}}{\rm M_{\odot }\,yr^{-1}},
\label{Rwind}
\end{equation}
 where $\eta$ = 0.25 ($\eta$ = 0.50, 0.75, and 1.00 are discussed in Section\,\ref{discussion}), and $R_{1}$, $L_{1}$ and $ M_{1}$ are the radius, luminosity and mass of the donor, respectively. However, for the AGB,  the formulation of \citet{1993ApJVassiliadis}  
\begin{equation}
{\rm log} \dot{M}_{\rm VW} = -11.4 + 0.0125\,[P_{0} - 100\,{\rm max}(M_{1} - 2.5, 0.0)],
\end{equation}
where $P_{0}$ is the Mira pulsation period.

 The mean accretion rate of the accretor: $\dot{M}_{2a}= {\rm min} \left \{ 0.8\left | \dot{M}_{\rm 1w} \right | , \dot{M_{2}}   \right \}$, where $\dot{M}_{\rm 1w}$ is the mass-loss rate of the wind from the donor, and $\dot{M_{2}}$ is got from \citet{1944MNRASBondi}
\begin{equation}
 \dot{M_{2}}= \frac{1}{\sqrt{1-e^{2}}}\left ( \frac{{\rm G\,}M_{2}}{v_{\rm w}^{2}} \right )^{2}\frac{\alpha_{\rm acc} }{2\,a^{2}}\frac{1}{\left (1+\frac{{\rm G}\,\left ( M_{1}+M_{2} \right )}{a\,v_{\rm w}^{2}}\right )^{\frac{3}{2}}}\left | \dot{M}_{\rm 1w} \right|,
\end{equation}
where $v_{\rm w}^{2}= 2\,\beta _{\rm w}\frac{{\rm G}\,M_{1}}{R_{1}}$, $M_{2}$ is the mass of the accretor. We choose $\beta _{\rm w} = 1/8 $, and $\alpha_{\rm acc} = 1.5$, as suggested in the \citet{2002MNRASHurley}. In this paper, the criterion of wind accretion channel in our computations is that the binary system does not undergo Roche lobe overflow, and the mass of the accretor is $10^{-10}\, M_{\odot}$ larger than its initial mass because of the wind accretion.

\subsection{The initial parameters for Monte Carlo simulation}

In the study of BPS, some initial parameters are required to be input in the Monte Carlo simulation (the sets with different parameters in this work are shown in Table\,\ref{tab:models}):  

(1) The primary mass is generated based on the initial mass function (IMF) of \citet{1979ApJSMiller}
\begin{equation}
  M_{1}^{\rm i} = \frac{0.19\,X}{(1-X)^{0.75}}+0.032\,(1-X)^{\frac{1}{4}},
\end{equation}
where a random number $ X$ is uniformly distributed in the range of $0 \sim 1$, and $M_{1}^{\rm i}$ is the initial mass of the primary, which is between $0.1\,\rm{ M_{\odot}} $ and $100\,\rm {M_{\odot}}$. 
 
(2) The initial distribution of mass ratio $q' = M_{2}^{\rm i} / M_{1}^{\rm i}$ is assumed to be

     (i) $n\,(q') = 1,$  $0 < q' \leq 1$, a constant mass ratio distribution \citep{1992ApJMazeh}.

     (ii)  $ n\,(q') = 2q' $, a rising distribution.

     (iii)  uncorrelated, two components are chosen randomly and independently from the same IMF. 

(3) The distribution of initial separation or initial orbital period is assumed to be

       (i) a power-law distribution for close separation, and a uniform distribution in ${\rm log}\,a$ for wide systems \citep{1995MNRASHanb}
\begin{equation}
    a\cdot n\,(a) = \begin{Bmatrix}
           0.07\,(a/a_{0})^{1.2},& a\leq a_{0},\\ 
           0.07 ,  & a_{0} <  a  < a_{1}, 
           \end{Bmatrix} 
\end{equation}
where $a_{0} = 10\,R_{\odot}$, and $a_{1} = 5.75 \times 10^{6}\,R_{\odot}$. Around half of the systems with initial orbital periods less than 100\,yr are considered as the binary systems. 

(ii)  the binary systems occupying half of samples follow a roughly log-normal Gaussian distribution with a mean of ${\rm log}P = 5.03$ and $ \sigma _{{\rm log} P} = 2.28 $ \citep{2010ApJSRaghavan}, where $P$ is in days.

\subsection{The CMD of standard model}
\label{sec:2.3}
We obtained the simulated photometry ($\rm G_{BP}-G_{RP}$ color and G-band magnitude) with GAIA filters (GAIA EDR3) using the PARSEC Bolometric Correction database \citep{Chen2019A&A}. It is difficult to distinguish two components of a binary system at the distance of GCs for photometric observations. Instead, the combined luminosity of the binary system is observed. Thus, we consider the sum of the luminosities of the two components as the total light of the binary system. Combining with the definition of the magnitude, we can get the synthetic magnitude of the two components with the following formula, already given by \cite{2015ApJXin}:
\begin{equation}
M_{\rm i}= M_{\rm i,1} - 2.5 \times {\rm log}\,(1 + 10^{\frac{M_{\rm i,1} - M_{\rm i,2}}{2.5}}),
\end{equation}
where $ M_{\rm i,1} $ and $M_{\rm i,2} $ represent the i-band (i can be $\rm G$, $\rm G_{BP}$ or $\rm G_{RP}$) magnitudes of the primary star and the secondary star of the binary system, respectively. $ M_{\rm i} $ is the synthetic magnitude of the binary system.

\begin{figure*}
\begin{center}
\includegraphics[width=0.7\textwidth]{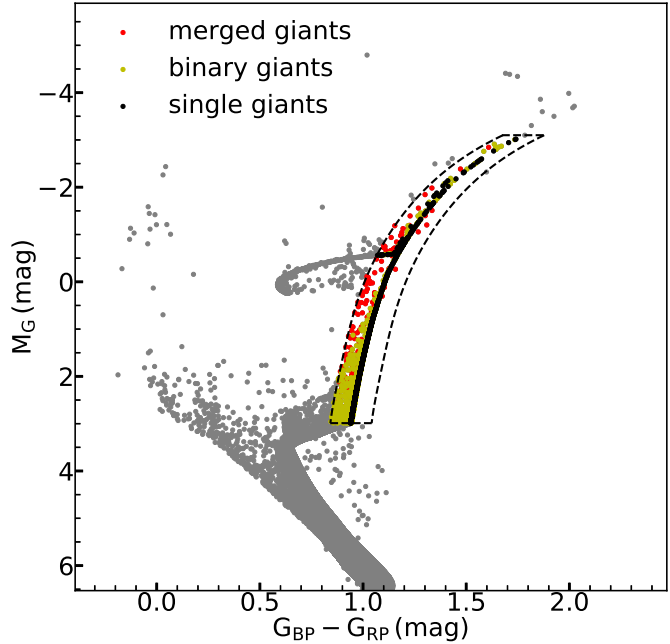}
\caption{CMD of all the systems (gray dots) at the age of 12\,Gyr for the standard model (set\,1). The black dashed lines mark the boundaries of the RGB of this simulated GC, and the RGB width is 0.2\,mag. The black dots represent single giants. The yellow dots and red dots are synthetic binary giants and merged giants produced from binary interactions, respectively.}
\label{fig:Giant}
  \end{center}
\end{figure*}

Fig.\,\ref{fig:Giant} shows the CMD of all the systems at the age of 12\,Gyr for the standard model (set\,1). The RGB widths are different in various GCs, and they are in a range of $0.2 \sim 0.4$\,mag \citep{2017MNRASMilone}. The black dashed lines mark the boundaries of the RGB of this simulated GC, and the RGB width is 0.2\,mag centered on the first giant branch (GB) of simulated single stars. For simplicity, all the samples in the box with the black dashed lines are considered as giants in this paper, although these samples in the box actually are consisted of Hertzsprung Gap (HG) stars, first giant branch (GB) stars, core helium burning (CHeB) stars, asymptotic giant branch (AGB) stars and even binaries. The post-BI stars in the box are called post-BI giants, which include binary giants (with two components) and merged giants produced from binary interactions. Giants from evolution of the primordial single stars here after are referred to as the single giants in this paper.

\begin{table}
  \centering
  \caption{Parameters used to produce samples. n\,($q$) and n\,(${\rm log}_{10}a)$ are initial mass ratio distribution and separation distribution, respectively.} 
    \label{tab:models}
  \begin{tabular}{lcccc} % three columns, alignment for each
     \hline
     \hline
    set  & n\,($q$)  & [Fe/H] & n(${\rm log}_{10}a$)  \\
    \hline  
    1 (standard)& constant   & -1.3   & exponential + constant \\ 
    2 & 2$q$         & -1.3   & exponential + constant  \\
    3 & uncorrelated    & -1.3   &exponential + constant \\ 
    4 & constant   & -1.3   & Gaussian \\
    5 & constant   & -2.3  & exponential + constant  \\
    6 & constant   & -1.9    & exponential + constant \\
    7 & constant   & -1.6    & exponential + constant\\
    8 & constant   & -0.7    & exponential + constant \\    
    9 & constant   & -0.3    & exponential + constant \\
    10 & constant   & 0.0    & exponential + constant\\

    \hline
  \end{tabular}
\end{table}

\section{Results}
\label{result}

%\subsection{The results of standard model (model A)}
\subsection{The fractions and distributions of post-BI giants}

\begin{figure*}
%\flushleft
\includegraphics[width=1.0\textwidth]{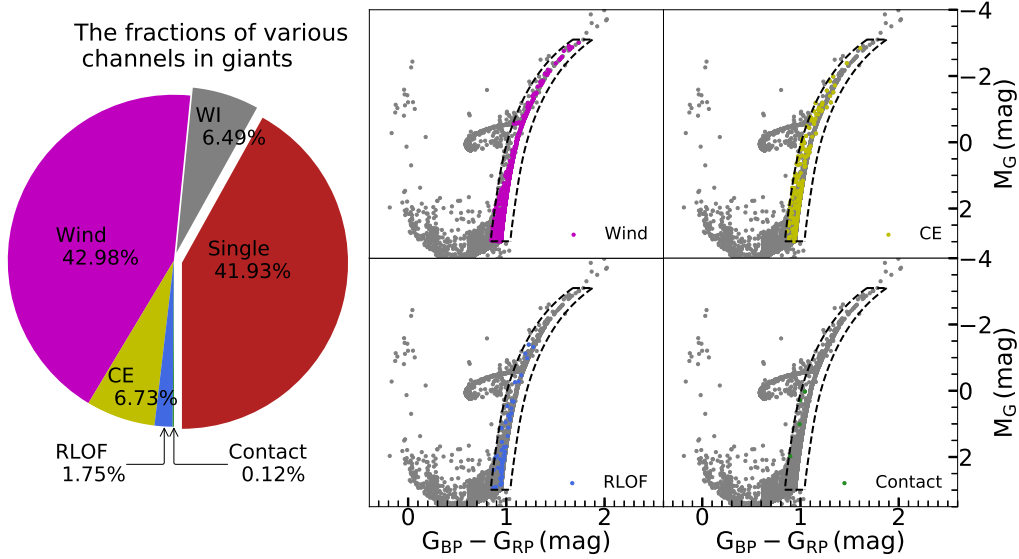}
\caption{The pie chart on the left indicates the fractions of post-BI giants for the simulation set 1 at the age of 12\,Gyr from four channels: RLOF channel (RLOF), common envelope channel (CE), contact binary channel (Contact) and wind accretion channel (Wind). The fractions of giants from binaries without interactions (WI) and single giants (Single) are also shown in this pie chart. The four panels on the right show the distributions of post-BI giants from four channels in CMDs.}
\label{fig:RGB}
\end{figure*}

The post-BI giants from various channels may have different properties. Fig.\,\ref{fig:RGB} shows their fractions and distributions in CMDs for the simulation set\,1 at the age of 12\,Gyr. We find that nearly half of all the giants ($ \sim 53.06\%$) have two components. As shown in the left pie chart in Fig.\,\ref{fig:RGB}, almost half of the giants ($\sim$ 51.58\%) underwent the binary interactions. The significant proportion (about 42.98\%) of the giants evolved from the wind accretion channel. The giants produced from the common envelope channel account for approximately 6.73\% of the whole giant samples, which is nearly four times the fraction of the giants from the RLOF channel ($\sim$ 1.75\%). The giants produced from the contact binary channel only may be of the order of 0.12\% of all the giants. Around 6.49\% of the giants are indeed in binary systems initially, but they do not experience any binary interactions and evolve similar to single stars. Besides, the fraction of single giants is $\sim$ 41.93\%. The right panels in Fig.\,\ref{fig:RGB} present the distributions of the giants from four binary-interaction channels (RLOF, CE, Contact and Wind) in CMDs. Obviously, the existence of post-BI giants can lead to the broadening of the RGB, although many of them occupy positions similar to single giants in the CMD.

%have undergone

\subsection{The fractions and distributions of E-BSSs}

\begin{figure*}
\centering
\includegraphics[width=1.0\textwidth]{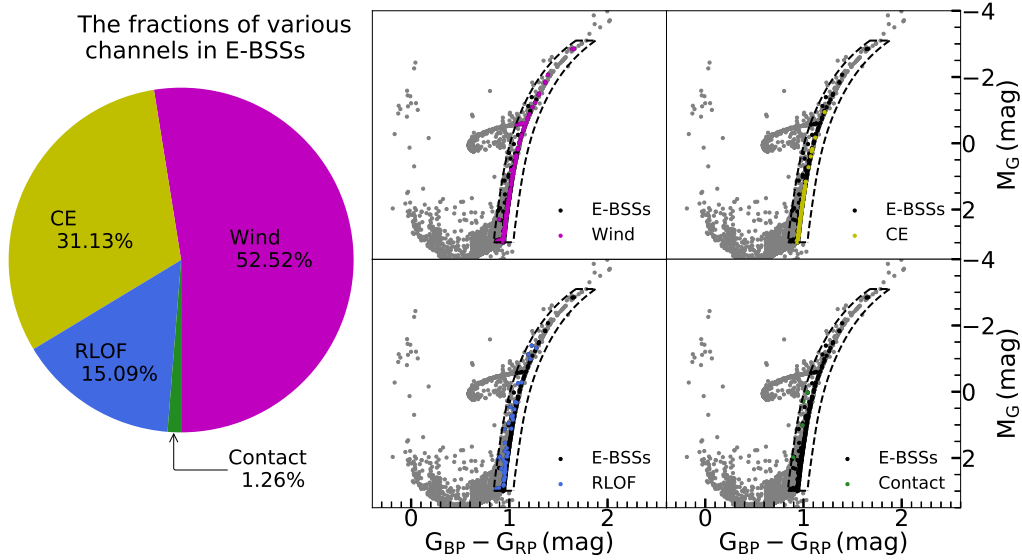}
\caption{The pie chart on the left indicates the fractions of the E-BSSs from four channels in all E-BSSs for the simulation set 1 at the age of 12\,Gyr. The four panels on the right show the distributions of E-BSSs from four channels in CMDs. The black samples in the four panels on the right present all the E-BSSs.}

\label{fig:BSS}

\end{figure*}
As a kind of special product from binary interactions, E-BSSs are suitable examples to explore the effects of such interactions on the giants. In our models, the MS lifetime of a BSS is older than that of a standard single star with the same mass, and it is named as E-BSS when it evolves to the post-MS phases (e.g. on the RGB). All of the binary-interaction channels above can generate E-BSSs. Fig.\,\ref{fig:BSS} shows the fractions of E-BSSs from various channels and their distributions in CMDs for the simulation set 1. According to our simulations, the E-BSSs occupy around 9.60\% of the giants. The left pie chart in Fig.\,\ref{fig:BSS} shows the fractions of E-BSSs from various channels at 12\,Gyr. Approximately half of the E-BSSs ($\sim$ 52.52\%) are produced from the wind accretion channel. The E-BSSs formed from the common envelope channel make up $\sim$31.13\%, which is around two times the fraction of the E-BSSs formed from the RLOF channel ($\sim 15.09\%$). About 1.26\% of E-BSSs come from the contact binary channel. The right panels in  Fig.\,\ref{fig:BSS} represent the distributions of the E-BSSs from four channels of binary interactions in CMDs. Most of the E-BSSs mix with the single giants, compared with the distribution of the single giants in Fig.\,\ref{fig:Giant}.

\subsection{The mass distributions of giants and E-BSSs}

\begin{figure*}

\includegraphics[width=1\textwidth]{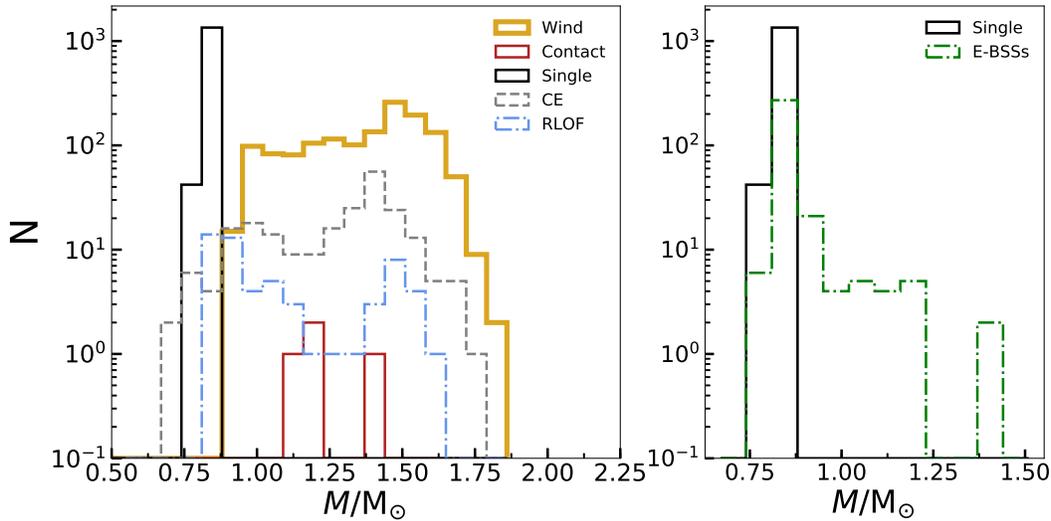}
\caption{Distribution of giant masses for the simulation set 1 at 12\,Gyr. The left panel shows the mass distribution of single giants and binary systems (the total mass of the two components in a system). The right panel shows the mass distribution of of E-BSSs and single giants. }
\label{fig:mass profile}
\end{figure*}

The masses of post-BI giants are expected to be different from that of single giants. Fig.\,\ref{fig:mass profile} shows the mass distribution of the giant samples for the simulation set\,1 at 12\,Gyr. The left panel shows the comparison of the mass distribution of single giants and that of the post-BI systems (e.g. the mass of the merged giant or the total mass of the post-BI binary system). The masses of single giants are around 0.80\,$\rm M_{\odot}$, which is close to the mass of the MSTO star in this simulated cluster. The masses of most post-BI systems are larger than that of the single giants, which suggests that the post-BI systems may sink into the center of the cluster due to mass segregation. In the right panel, we compare the mass distribution of E-BSSs and that of single giants. It is clear that most of the E-BSSs are massive than the single giants, and some of them even have a mass of nearly 1.4\,$\rm M_{\odot}$, which are merged giants from the contact binary channel or common envelope channel.

\subsection{The initial mass ratio distribution of post-BI giants from four channels}

\begin{figure}
\begin{center}
\includegraphics[width=0.7\textwidth]{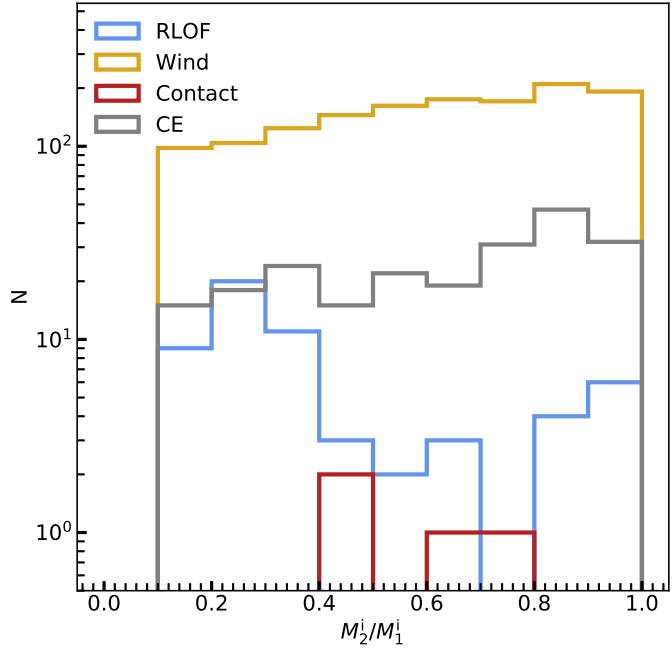}
\caption{The initial mass ratio distribution of post-BI giants from four channels for the simulation set 1 at 12\,Gyr.}
\label{fig:mass_ratio}
   \end{center}
\end{figure}

The initial mass ratio ($q' = M_{2}^{\rm i}/M_{1}^{\rm i} $) plays an important role in the binary evolution, which especially affects the stability of mass transfer. Fig.\,\ref{fig:mass_ratio} shows the initial mass ratio distribution of post-BI giants from different channels. The number of post-BI giants from the wind accretion channel increases with the increasing initial mass ratio. The reason for this increasing trend is that the majority of the accretors are required to have a initial mass slightly smaller than that of the MSTO stars (e.g. $\sim \rm 0.8\,M_{\odot}$) due to the low accretion rate in the wind accretion channel. At the same time, the number of the donors decrease with the initial mass as the result of IMF. Therefore, the post-BI giants in binaries with higher initial mass ratio are more than that with lower mass ratio. Because of the similar reason, the number of post-BI giants in the CE channel also slightly increases with the increasing initial mass ratio. For the RLOF channel, the majority of post-BI giants with $0 < q' = M_{2}^{\rm i}/M_{1}^{\rm i} < 0.4 $ experienced the unstable mass transfer, while those with $q'>0.4$ experienced stable RLOF mass transfer, according to the critical mass ratio ($ q'_{\rm c} = M_{2}/M_{1}= 1/3 $) for the MS donor in Section\,\ref{RLOF} . Moreover, post-BI giants from the contact binary channel can only be formed by stable RLOF mass transfer. Therefore, they have a mass ratio larger than 0.4.

 %There is an stable RLOF mass transfer case in the post-BI giants from RLOF channel with $0 < q' = M_{2}^{\rm i}/M_{1}^{\rm i} < 0.4$. It is because the donor in this system is a Hertzsprung Gap star, which has a different critical mass ratio ($ q'_{\rm c} = M_{2}/M_{1}= 1/4 $ for Hertzsprung Gap donor).

%The post-BI giants from RLOF channel with $ q' = M_{2}^{\rm i}/M_{1}^{\rm i}> 0.4$ experienced stable RLOF mass transfer, and all of post-BI giants from the contact binary channel underwent the stable RLOF mass transfer. It is necessary for CE channel to have a giant or giant-like donor and $q' < q'_{\rm c}=M_{2}/M_{1}$. The critical mass ratio relates to the giant's core mass and radius. The gray line in Fig.\,\ref{fig:mass_ratio} indicates that the binary systems with various initial mass ratios can undergo the common envelope channel.}

\subsection{The evolutionary type of the donor at the beginning of binary interactions }

\begin{figure}
\begin{center}
\includegraphics[width=0.7\textwidth]{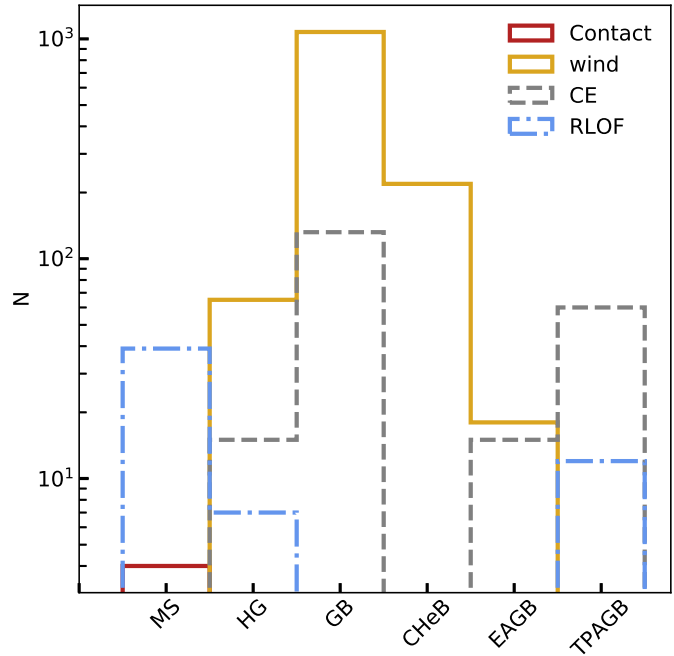}
\caption{Distribution of the evolutionary type of the donor at the moment when the donor fills its Roche lobe (for CE, Contact and RLOF channel) or the accretor begins to accrete material (for the wind accretion channel). MS, HG, GB, CHeB, EAGB, and TPAGB indicate main sequence, Hertzsprung Gap, first giant branch, core helium burning, early asymptotic giant branch and thermally pulsing AGB, respectively. }
\label{fig:type}
   \end{center}
\end{figure}

The post-BI stars may have surface abundance anomalies because of binary interactions, e.g. surface pollution by mass transfer, or rotational mixing in rapidly rotating mergers. The evolutionary phase of the donor (as the source of polluted material) may be one of the main parameter affecting the surface abundance anomalies of the accretor. Fig.\,\ref{fig:type} presents the distribution of the evolutionary type of donor at the beginning of binary interactions. When the donors are on MS phase, most of the binaries experience RLOF channel and the minority of them undergo the contact binary channel. A part of the binaries experience the RLOF channel, and most of them are in the common envelope, when the donors are on HG or thermally pulsing AGB (TPAGB). The common envelope occurs as long as the donors are post-MS stars (e.g. HG, GB, EAGB and TPAGB), but except CHeB. That is because the CHeB stars have extreme small radius and it is difficult to overflow the Roche lobe. The wind accretion occurs when the donors are on the HG, GB, CHeB, and EAGB.

\subsection{The influence of the model parameters}

\begin{figure}
\begin{center}
\includegraphics[width=0.7\textwidth]{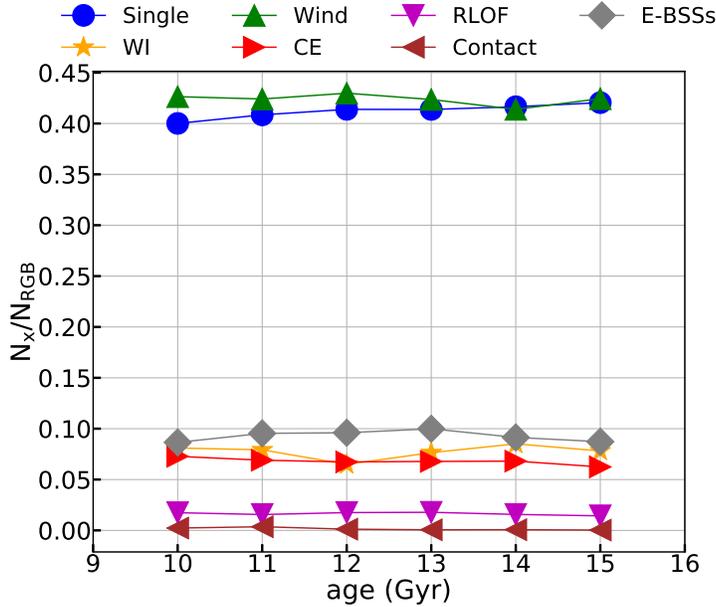}
\caption{The fractions of giants produced from different channels as a function of age for the simulation set 1. These channels include: single giants (Single), giants in binary but without interaction (WI), RLOF channel (RLOF), common envelope channel (CE), contact binary channel (Contact), wind accretion channel (Wind) and evolved blue straggler stars (E-BSSs).}
\label{fig:X age}
   \end{center}
\end{figure}

\begin{figure}
\begin{center}
\includegraphics[width=0.7\textwidth]{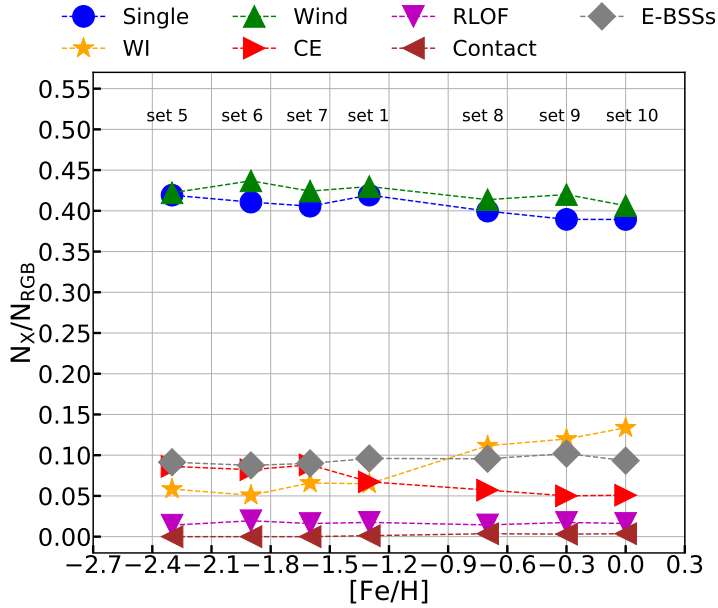}
\caption{Dependence of the fractions of giants produced from different channels on the metallicities at 12\,Gyr.}
\label{fig:X z}
   \end{center}
\end{figure}

The model parameters of the Monte Carlo simulation may play an important role in the fractions of post-BI giants and E-BSSs. An interesting question is whether these fractions depend on age. Fig.\,\ref{fig:X age} shows the fractions of giants produced from different channels versus age. It seems that the fractions of post-BI giants in these channels do not show significant changes over time. Moreover, the fraction of E-BSSs in giants is also roughly constant, which shows that about 10\% of the giants are E-BSSs. In order to explore the effect of the metallicity on the results, we show the dependence of the fractions of post-BI giants on metallicities in Fig.\,\ref{fig:X z}. These fractions do not strongly depend on the metallicity. Furthermore, E-BSSs in giants show no dependence of the fraction on the metallicity, and they can account for about 10\% of the giants in different metallicity environments.

\begin{figure}
\begin{center}
\includegraphics[width=0.7\textwidth]{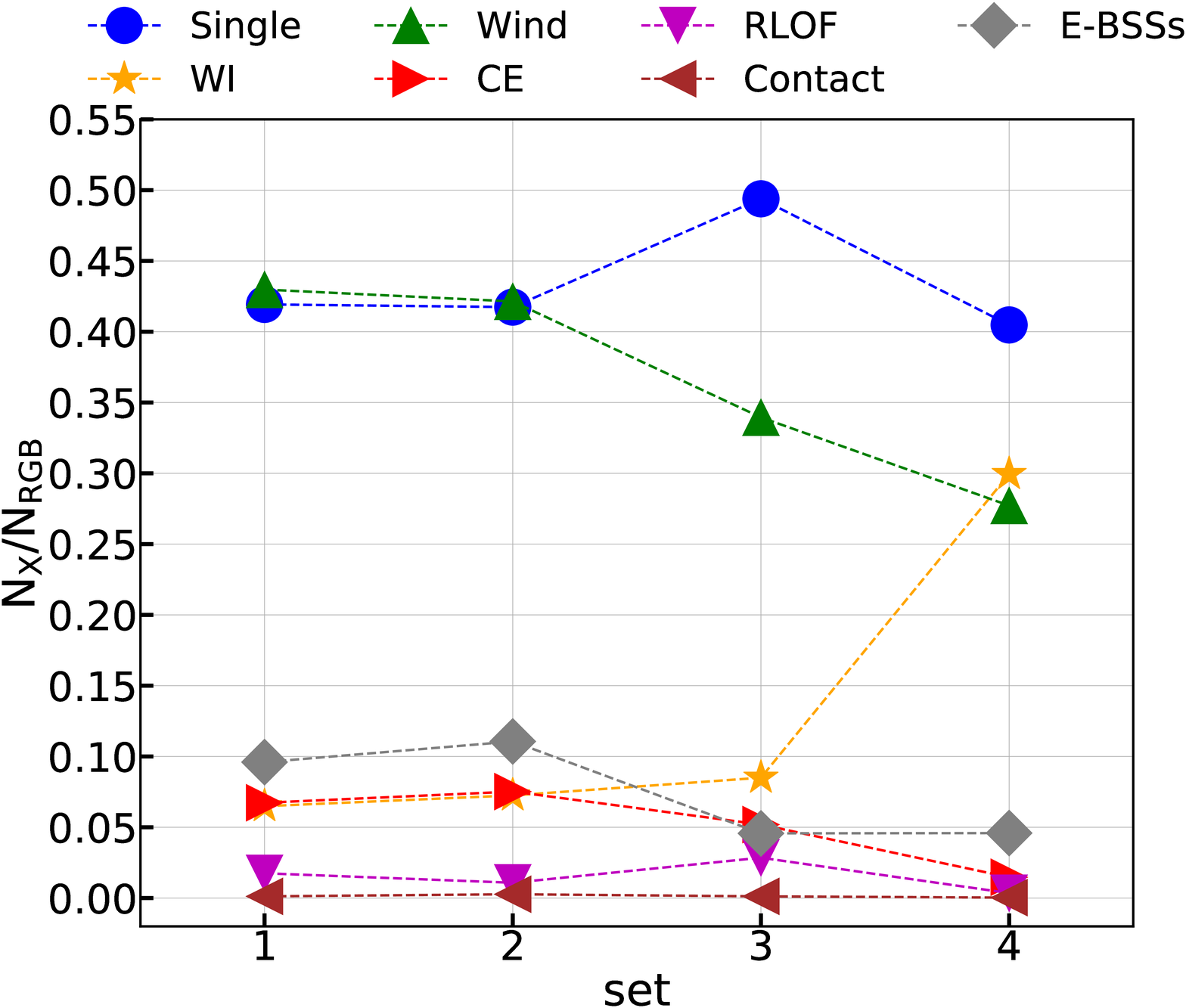}
\caption{Dependence of the fractions of giants produced from different channels on the initial parameters at 12\,Gyr.}
\label{fig:model}
   \end{center}
\end{figure}

We also investigate the effects of initial distributions of the mass ratio and separation (or orbital period) on our results in Fig.\,\ref{fig:model}. The sets\,1, 2 and 3 are simulations with different initial distributions of mass ratio. The results from set\,1 are similar to that from set\,2. However, the set\,3 shows relatively high fraction of single giants. That is because the initial masses of most accretors are significantly smaller than 0.8\,$\rm M_{\odot}$ in set\,3, which can not appear on the RGB at 12\,Gyr even through binary interactions. Thus, the fraction of giants produced from binary interactions is lower, and the fraction of single giants is relatively higher in set\,3, compared with the set\,1. The set\,4 with initial orbital period of Gaussian distribution generates more giants which do not undergo the binary interactions, due to the larger initial orbital period in the binary samples. The sets\,3 and set\,4 reveal that about 4.59\% of giants are E-BSSs.

\section{Discussion and conclusions}
\label{discussion}
Binary interactions have been suggested as one of the natural ways to form the chemically peculiar stars on the RGB. However, the fraction of post-BI stars on the RGB is still unclear. In order to well understand the possible contribution of the binary interactions to the multiple stellar populations, we investigate the characteristics of post-BI giants by employing the BPS method. We find that around 51.58\% of giants underwent the binary interactions in the simulation set\,1, assuming an initial binary fraction of nearly 50\%. A large fraction ($\sim$ 42.98\%) of giants formed from the wind accretion channel. 

Besides, in the simulation set\,1, nearly 53.06\% of giants belong to binary systems with two components, which may be observed as binary systems. The binary fraction in RGB stars is different in various GCs. \citet{2015A&ALucatello} found a binary fraction of $2.2\% \pm 0.5\%$ using a sample of ten GCs, and \citet{2018ApJDalessandro} indicated the binary fraction in NGC\,6362 is $\sim 9.0\%$. These observed binary fractions are smaller than our results. Such differences can be explained by dynamical interactions or a lower initial fraction of binaries in GCs. Some works studied dynamical evolution of GCs, including the effects of dynamical interactions on the binaries in GCs \citep{2005MNRASIvanova,2010ApJChatterjee,2015MNRASHong,2016MNRASHong,2016MNRASWang}. According to the N-body simulations from \citet{2005MNRASIvanova}, the dynamical interactions reduce the binary fraction in the core of a given GC (14\,Gyr) to 7\%, from the initial binary fraction of 50\%. Therefore, the binary fraction in giants will be far lower than our theoretical estimates, if the dynamical interactions (e.g. dynamical disruptions) are considered. In addition, some post-BI giants may also not be observed as binary systems, if they are mergerd giants or have a very faint white-dwarf companion.

%Some of the giants in GCs indeed underwent binary interactions (post-BI giants), even though they are not observed as binary systems now, due to the dynamical disruptions. }

%may decrease the binary fraction as given by N-body simulations \citep{2015MNRASHong,2016MNRASHong}.}

We study the effects of main model parameters (age, metallicity, and initial distribution of binaries) on the properties of giants in a simulated GC. We find that the initial distributions of separation (or orbital period) and mass ratio are important parameters remarkably affecting the fractions of post-BI giants. However, our results  depend weakly on the initial metallicity and age from our simulations. The reason for the stable pattern is that the numbers of giants from various channels show the similar variation trend: the number of giants from various channels is decreasing with the increasing age, and is increasing with the increasing metallicities. The post-BI giants show different properties compared with the single giants. They should have a larger mass than single giants, similar to E-BSSs \citep{2016ApJFerraro,2016ApJParada}, due to mass transfer or merger in the previous binary systems.

E-BSSs are known to be possible candidates of post-BI stars on the RGB of GCs. Recently, \citet{2019A&ARvGratton} roughly estimated the fraction of E-BSSs in giants is 6\%. We find that E-BSSs account for about 10\% of the RGB stars in the simulation set\,1, and it is larger than the fraction given by \citet{2019A&ARvGratton}, due to the different distribution of initial orbital period. We get the similar fraction when we adopt the initial orbital period of Gaussian distribution in set\,4. Furthermore, \citet{2020ApJSun} found a large fraction of the “young” $[\alpha/{\rm Fe}]$-enhanced stars in their red-clump samples possibly including the contribution of stars from the field and GCs, and this intriguing population might be from the binary process such like E-BSSs. Our results of simulated GCs (e.g. simulation set\,1) show that the fraction of E-BSSs could be relatively high, which is consistent with the results in Milky Way in \citet{2020ApJSun}. So our work may be helpful to further understand the Galactic chemical evolution.

%------------------------------------------------------------------------------------

There are some uncertain parameters ($\lambda$ in Equation\,\ref{CE} and $\eta$ in Equation\,\ref{Rwind}) in our computations. We test the simulations set\,1 with different uncertain parameters ($\lambda$ = 0.25, 0.5, 0.75, 1.00; $\eta$ = 0.25, 0.50, 0.75, 1.00). The fraction of E-BSSs slightly increases with the increasing $\lambda$, from 8.48\% for $\lambda$ = 0.25 to 10.86\% for $\lambda$ = 1.00. That is because $\lambda$ is related to the number of giants produced from the common envelope channel, which affects the number of E-BSSs. We also find that the fraction of giants formed from the wind accretion channel slightly increases with the increasing $\eta$. The fraction of giants formed in this channel is 42.98\%, 44.22\%, 46.16\%, and 47.24\%, for $\eta$ = 0.25, 0.50, 0.75 and 1.00, respectively.

%------------------------------------------------------------------------------------
Obviously, the post-BI giants are likely to show anomalous abundances, compared with the single giants. However, these anomalous abundances may depend on the abundances of material transferred from the donors and the dilution on the surface of the accretors \citep[e.g.][]{2020MNRASWei}. For example, post-BI giants having accreted from a TPAGB star may be characterized by an overabundance of s-process elements, such as Ba stars \citep{1995MNRASHana}, while post-BI giants having accreted previously from a HG star may show the nitrogen enrichment, and the carbon depletion \citep{2020MNRASWei}, similar to the chemical anomalies observed in GCs. Our results show that the accreted material of the post-BI giants may come from different donor types (from MS to TPAGB) by different channels of binary interactions (e.g wind accretion, RLOF). But it is still uncertain which type of post-BI giants may have anomalous abundances similar to the observation of GCs. More works are still required to confirm the contribution of post-BI giants to the observed chemically peculiar population in GCs.

\begin{acknowledgements}
We thank Hongwei Ge, Yangyang Zhang and Heran Xiong for their helpful discussions about the BPS method, and we also thank Jiangdan Li, Jiao Li, Gaurav Singh and Carine Babusiaux for their precious suggestions about the observations. We greatly thank Yan Gao for the language of this paper. We are grateful to the National Natural Science Foundation of China (Grant No. 12073070, 11733008, 11873085, 11521303, 12073071, 11873016, 11903075, 12003027, 11973081), the Natural Science Foundation of Yunnan Province (No. 2017HC018, 202001AT070058, 202001AU070054), Youth Innovation Promotion Association of Chinese Academy of Sciences (Grant No. 2018076, 2012048), and the Chinese Academy of Sciences (CAS; KJZD-EW-M06-01) for support. This work is supported by CAS `Light of West China' Program. This work is also supported by National Key Basic R\&D Program of China via 2019YFA0405500, the LAMOST Fellow project, funded by China Postdoctoral Science Foundation (Grant No. 2019M653504, 2020T130563), Yunnan province postdoctoral Directed culture Foundation, and the Cultivation Project for LAMOST Scientic Payoff and Research Achievement of CAMS-CAS.

\end{acknowledgements}

\bibliographystyle{raa} % style aa.bst
\bibliography{RAA-2021-0068}

\end{document}